\documentclass{emulateapj}
\usepackage{epsfig}

\newcommand{\saxj}{\mbox{SAX J1808.4$-$3658}}
\newcommand{\rxte}{\textit{RXTE}}
\newcommand{\us}{\mu s}
\newcommand{\uHz}{\mu\rm\,Hz}

\newcommand{\hzs}{\rm\,Hz\,s^{-1}}
\newcommand{\msun}{\,M_{\odot}}
\newcommand{\nudot}{\dot{\nu}}

\begin{document}

\title{Accelerated Orbital Expansion And Secular Spin-Down of the Accreting Millisecond Pulsar \saxj\ }
\shorttitle{Evolution of \saxj}
\shortauthors{Patruno et~al.}

\author{
  Alessandro Patruno\altaffilmark{1},
  Peter Bult\altaffilmark{1},
  Achamveedu Gopakumar\altaffilmark{2}
  Jacob M. Hartman\altaffilmark{3,4},
  Rudy Wijnands\altaffilmark{1} ,
  Michiel van der Klis\altaffilmark{1},
  Deepto Chakrabarty\altaffilmark{5}}
\altaffiltext{1}{Astronomical Institute ``Anton Pannekoek,''
  University of Amsterdam, Science Park 904, 1098 XH Amsterdam, The Netherlands}
\altaffiltext{2}{Department of Astronomy and Astrophysics,
Tata Institute of Fundamental Research, Colaba, Mumbai-400 005, India}
\altaffiltext{3}{NASA Postdoctoral Program fellow}
\altaffiltext{4}{Jet Propulsion Laboratory, Pasadena, CA 91109, USA}
\altaffiltext{5}{Department of Physics and Kavli Institute for Astrophysics
  and Space Research, Massachusetts Institute of Technology, Cambridge, MA
  02139, USA}

\begin{abstract}
The accreting millisecond pulsar \saxj\ has shown a peculiar orbital
evolution in the past with an orbital expansion much faster than
expected from standard binary evolutionary scenarios. Previous limits
on the pulsar spin frequency derivative during transient accretion
outbursts were smaller than predicted by standard magnetic accretion
torque theory, while the spin evolution between outbursts was
consistent with magnetic dipole spin-down. In this paper we present
the results of a coherent timing analysis of the 2011 outburst
observed by the \textit{Rossi X-ray Timing Explorer} and extend our
previous long-term measurements of the orbital and spin evolution over
a baseline of thirteen years. We find that the expansion of the 2 hr
orbit is accelerating at a rate
$\ddot{P}_b\simeq1.6\times10^{-20}\rm\,s\,s^{-2}$ and we interpret
this as the effect of short-term angular momentum exchange between the
mass donor and the orbit.  The gravitational quadrupole coupling due
to variations in the oblateness of the companion can be a viable
mechanism for explaining the observations.  No significant spin
frequency derivatives are detected during the 2011 outburst
($|\nudot|\lesssim4\times10^{-13}\hzs$) and the long term spin-down remains
stable over thirteen years with $\nudot\simeq -10^{-15}\hzs$.
\end{abstract}


\section{Introduction}

The transient X-ray binary \saxj\ is the first accreting millisecond
X-ray pulsar (AMXP) discovered \citep{wij98} among the 14 systems
currently known.  It is also the best sampled AMXP thanks to its
relatively short recurrence time (1.6-3.3 yr) and the continuous
coverage of the \textit{Rossi X-ray Timing Explorer} (\rxte\ ) which
has extensively monitored all outbursts since 1998. The presence of
pulsations reveal the spin of the accreting pulsar ($\approx
401\rm\,Hz$) and allow the study of torques that act upon the neutron
star.  This is particularly valuable because the spin evolution
reveals details of the recycling mechanism that transforms a newly
born slowly rotating neutron star into a millisecond pulsar via
accretion \citep{alp82, rad82}.  In our previous works
(\citealt{har08,har09}) we reported upper limits on the spin-up of the
pulsar due to accretion torques (during outbursts) with a pulsar spin
frequency derivative of $|\nudot|\lesssim 2.5\times10^{-14}\hzs$.
This value is smaller than predicted by accretion theory
(\citealt{gho79}) if the spin frequency of SAX J1808.4-3658
substantially differs from the equilibrium spin frequency at the
accretion rate close to the peak of the outburst.  The long term spin
evolution of the pulsar reveals a constant spin-down of magnitude
$-5.5\pm1.2\times10^{-16}\rm\,Hz\,s^{-1}$ possibly due to
magnetic-dipole radiation acting during quiescence\footnotemark[6]\footnotetext[6]{In the
  abstract of \citealt{har09} a wrong value of
  $-5.5\pm1.2\times10^{-18}\rm\,Hz\,s^{-1}$ is reported, due to a
  typographical error.}, for a surface magnetic field of the pulsar
$B\simeq1.5\times10^8$G, in line with the expected field strength of
millisecond radio pulsars.

\saxj\ is undergoing an unexpectedly fast orbital evolution with the
orbital period increasing on a timescale of $\approx70$Myr
(\citealt{har08},\citealt{dis08}). The binary has an orbital period of 2.01 hr
\citep{cha98} and the donor star is a $0.05-0.1\msun$ brown dwarf
\citep{bil01,del08} suggesting that the orbital evolution should be
dominated by angular momentum loss via gravitational waves and
possibly by magnetic braking~\citep{tau06}. The timescale of the orbital
evolution is, however, too fast to be explained with such a scenario,
and non-conservative processes with large mass loss from the system
have been invoked (\citealt{dis08}, \citealt{bur09}).
\citet{har08,har09} suggested instead that interchanges of angular
momentum between the companion and the orbit can dominate the
short-term orbital evolution as seen in several binary millisecond
pulsars \citep{arz94, nic00}.

On October 31 2011 \textit{Swift}-BAT detected a new outburst of
\saxj\ \citep{mar11,pap11}. This is the 7th outburst observed since its
discovery \citep{int98} and the 6th monitored with \rxte.  We present
a coherent pulsation analysis of the outburst and we complete the study of
the spin and orbital evolution of \saxj\ over a baseline of thirteen yr. 

\section{X-ray Observations and Coherent Analysis}

We use all \rxte\ Proportional Counter Array (PCA; \citealt{jah06})
public data for the 2011 outburst (Program-Id 96027). We construct the
2-16 keV X-ray light-curve with PCA Standard2 data averaging the flux
for each observation and normalizing it in Crab units (see for example
\citealt{van03}). One burst is detected at MJD 55873.9 and all data
with a flux more than twice the pre burst level are removed from the
light-curve.

For the timing analysis we use all photons (excluding the burst
interval) in the energy band $\approx2-16$ keV (5-37 absolute
channels) in Event $122\us$ mode. The data are barycentered with the FTOOL
\textit{faxbary} by using the optical position 
\citep{har08} and the JPL DE405 solar system ephemeris. We fold
$\sim500$s long data segments in pulse profiles of 32 bins, keeping
only those with signal-to-noise $>3.3\sigma$, giving $<1$ false pulse
detection for the entire outburst. The S/N is defined as the ratio between the
pulse amplitude and its $1\sigma$ statistical error. The folding
procedure uses the preliminary ephemeris reported in \citet{pap11}.
A fundamental ($\nu$) and a first overtone ($2\nu$) are detected
in the pulse profiles. The TOAs are measured separately for each
harmonic to avoid that pulse shape variability affects the fiducial
point defining the pulse TOA (see \citealt{har08} for details). 

To follow the evolution of the orbit and the pulsar spin we fit
separately the two sets of TOAs (fundamental and first overtone) with
the software TEMPO2 \citep{hob06}. 
The initial model used is a Keplerian circular orbit and a constant
pulse frequency. We then repeat the folding procedure with the
new timing solution until we reach convergence to the final orbital
and pulse parameters. To calculate the errors on the rotational
parameters we use Monte Carlo (MC) simulations, that account for the
presence of long-timescale correlations in the pulse TOAs
\citep{har08,arz94}. To verify the presence of a spin frequency
derivative we fit a pulse frequency and its time derivative to the
TOAs and run $10^4$ MC simulations to estimate the significance of the
measurements (see \citealt{har08} and \citet{pat09b} for further details).

\section{Results of the 2011 Outburst}

\subsection{X-ray Light-curve and Pulse Profiles}\label{pp}

\begin{figure}[t]
  \begin{center}
    \rotatebox{-90}{\includegraphics[width=1.1\columnwidth]{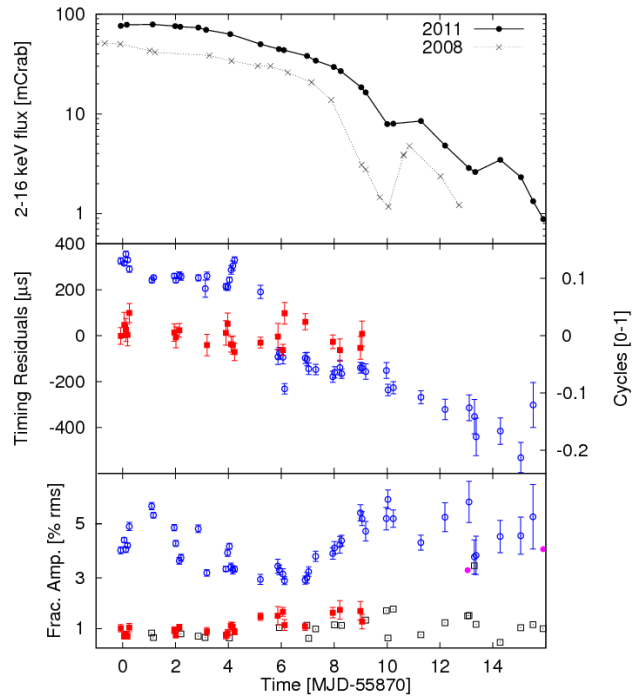}}
  \end{center}
  \caption{\textbf{Top panel:} X-ray light-curve (2-16 keV) of the
    2011 outburst (solid line) compared to the 2008 outburst (dotted
    curve). The data points are \rxte\ observations-long
    averages. \textbf{Middle Panel:} timing residuals for fundamental
    (blue dots) and first overtone (red squares) for a $\nu=const$
    model. Each data point is an \rxte\ orbit-long average. The
    fundamental has a jump of $\sim0.15$ at MJD $\approx55875$.
    \textbf{Bottom Panel:} Fractional amplitude of the fundamental
    (blue dots) and first overtone (red squares). The 95\% c.l. upper
    limits are indicated with pink circles (fundamental) and black
    open squares (first overtone).  \label{fig1}}
\end{figure}

The first \rxte\ pointed observation was taken on November 4,
$\approx5$ days after the beginning of the outburst \citep{mar11}.
The X-ray flux shows a peak at $\approx80$ mCrab, remarkably higher
than the peak luminosity in 2005 and 2008 and similar to the 1998 and
2002 values \citep{wij04}. Since the observations started 5 days
after the onset of the outburst, the true outburst peak at MJD 55868
(observed by \textit{Swift}-BAT) has been missed by \rxte\,. The
\rxte\ PCA light-curve of the 2011 outburst is shown in
Figure~\ref{fig1}.

The flux is observed to slowly decay over the entire duration of 
the main outburst, reaching a minimum
flux of 8 mCrab on MJD 55880. The flux then keeps decreasing with two
bumps at MJD 55881 and 55884, which might possibly be associated with
the flaring-tail stage, when quasi periodic bumps are
observed at low flux level (\citealt{wij01, har08, pat09c}).  However,
differently from the 2000 to 2008 outbursts, this phase is poorly
sampled and difficult to characterize. During the tail a strong 1
Hz modulation was reported for several outbursts (2000, 2002, 2005;
\citealt{van00, pat09c}) but it is not detected in any of the 2011
observations.

The source became undetectable on MJD 55885.9 (November 20) and the
monitoring abruptly ended on November 27, due to solar
constraints. These also prevented to establish whether at
the time of the last observation \saxj\ was in quiescence or in one of
the faint ($\sim10^{32}\rm\,erg\,s^{-1}$) states observed in the past
with \textit{Swift} and \textit{XMM-Newton}
observations\citep{wij03, cam08}. 

The pulse profiles are very sinusoidal 
until MJD 55874 and then change becoming skewed with a more
evident first overtone peaking on the right part of the
profile. At MJD 55880, during the possible flaring stage, the pulse
profiles become very sinusoidal again, a phenomenon never observed in
the previous outbursts.  This is reflected in the rms amplitude of the
two harmonics (see Fig~\ref{fig1}).


\subsection{Timing Noise and Error Estimation}\label{mc}

The timing solution of the 2011 outburst is displayed in Table~\ref{tab1} 
\begin{table}
\centering
\caption{\saxj\ Timing Solution for the 2011 Outburst }
\begin{tabular}{llll}
\hline
\hline
Parameter & Value & Stat. Error & Syst. Error\\
\hline
$\nu$ [Hz] & 400.97520981 & $7\times10^{-8}$ & $10^{-7}$\\
$|\nudot| [\hzs]$ & $<4\times10^{-13}$ & (95\% c.l.)&  \\
$P_b$ [s]  & 7249.162 & 0.003  & \\
$A_x sin(i)$ [lt-ms] & 62.798 & 0.005 &  \\ 
$T_{asc}$ [MJD] & 55896.895635 & 0.00002 & \\
$e$ &  $<10^{-4}$ &  (95\% c.l.) & \\
\hline
\end{tabular}\label{tab1}
\end{table}
while a previous analysis of the five outbursts observed with
\rxte\ is available in \citet{har08,har09}. The authors found strong
timing noise operating on the same timescales over which the pulse
frequency and its time derivative were measured. The 2011 TOA residuals
of a $\nu=const$ model, show the typical behavior 
observed in the previous outbursts in both harmonics.  

We detect a strong phase jump in the fundamental with magnitude of
0.15 cycles (0.3 ms) at MJD $\approx55874-55876$. This behaviour is
similar to what was previously seen during the 2002 and 2005 outbursts,
with phase jumps of $0.2$ cycles observed when the flux reached
the transition from slow to fast decay (\citealt{bur06},
\citealt{har08}, \citealt{pat09c}). The first overtone instead has no
phase jump but displays a short timescale (few minutes to
$\sim1$ day) scattering slightly in excess of that expected from
measurement errors alone.  

We use the phase information of the first overtone to phase connect
across the phase jump of the fundamental (see \citealt{har08} for
details of the procedure). A net spin frequency derivative (i.e.,
measured over the entire outburst length) is not detected, with upper
limits of $|\dot{\nu}|\lesssim 8.8\times10^{-13}\hzs$ at the 95\%
confidence level. By removing the TOAs of the first observation in
both harmonics (when timing noise is strong) and the 0.15 cycle phase
jump for the fundamental, we obtain similar results with a more
stringent constraint on the spin frequency derivative:
$|\dot{\nu}|\lesssim 4\times10^{-13}\hzs$ ($95\%$ c.l.).

The errors on the orbital parameters measured with the fundamental are
only marginally affected by timing noise since they are measured on
timescales (2 hr) different than the timing noise one ($\sim\,$days).
This is verified by calculating a power spectrum of the TOA residuals
and comparing the Poissonian level to the power at the orbital
frequency $1/P_b$. The excess power at $1/P_b$ is about $1.5$ and $3$
times the Poissonian level for fundamental and first overtone
respectively. We therefore rescale the statistical errors on the
orbital parameters by the same factor.

\subsection{X-ray Flux - Pulse Phase Correlation}\label{phaseflux}

In 2009 \citep{pat09d} proposed an alternative method to partially
account for the timing noise in the TOAs of AMXPs. X-ray flux
variations were found to be linearly correlated or anti-correlated
with the pulse phases. 
Instead of minimizing the rms of the TOA residuals, \citet{pat09d}
minimized the $\chi^2$ of a linear fit to the phase-flux correlation,
finding slightly different spin frequencies than those measured with
rms minimization methods. The reason of this difference is that
instead of treating timing noise as a red noise process of unknown
origin, the variations of the X-ray flux are assumed to
instantaneously affect the pulse phases. We found that while the
fundamental frequency always follows a correlation, the first overtone
in some cases behaves differently and we exclude it from our 2011
analysis. By repeating the same procedure outlined in \citet{pat09d}
for the 2011 outburst, we find a pulse frequency of
$\nu=400.97520981(7)\rm\,Hz$, where $1\sigma$ errors have been
rescaled by a factor 2.5 such that $\chi^2/dof=1$
(\citealt{bev03,pat09a}). The difference between this value and the
pulse frequency $\nu_{rms}$ obtained with standard rms minimization of
the TOA residuals is $\nu-\nu_{rms}=-0.15\pm0.08\uHz$.

\section{Results on The Long Term Evolution of SAX J1808.4-3658}

\subsection{Long Term Spin Frequency Evolution}

We first fit the change of the six constant pulse frequencies (from
1998 to 2011) of \saxj\ using the values reported in \citet{pat09d}
and the 2011 frequency obtained from the flux-phase correlation
technique. We rescale the errors of each spin frequency to give a
$\chi^2/dof=1$ and we fit a linear relation to the data. The fit
gives a $\chi^2=5.4$ for 4 dof, and a spin-down of
$\nudot=-1.65(20)\times10^{-15}\hzs$ (Fig.~\ref{fig2}). This is in
agreement with the value reported in \citet{pat09d}. A spin frequency
second derivative is not required by the fit, and we can place upper
limits of $|\ddot{\nu}|\lesssim 10^{-24}\rm\,Hz\,s^{-2}$ (95\% c.l.).

We also fit the 2011 outburst pulse frequency with the rms
minimization method (i.e., TEMPO2 plus MC errors), together with the
previous five spin frequency measurements obtained with the same
technique \citep{har09}. The $\chi^2$ is high, 19.5 for 4
dof and we ascribe this almost exclusively to the 2000
outburst spin frequency (see \citet{har09}).  If we
remove the 2000 data, the fit returns a
$\chi^2=3.57$ for 3 dof. The spin-down is constrained to be
$\nudot=-7.4(4)\times10^{-16}\hzs$. This estimate is within 2$\sigma$
from the value previously reported by \citet{har09} and within
1$\sigma$ when removing the 2000 outburst data. 

Since both fits of the long-term spin frequency evolution are
statitically acceptable, we cannot decide which of the two values
reported is closer to the true $\nudot$. The differences in the $\nu$
and long term $\nudot$ found with the rms minimization and with the
phase-flux correlation reflect a systematic uncertainty of
$\sim0.1\uHz$ that needs to be considered until the exact mechanism
behind timing noise is identified.

\subsection{Orbital Evolution}

To detect variations of the orbital period we use the procedure
explained in \citet{har08,har09} that requires estimates for the times
of passage through the ascending node $T_{asc}$ at various outbursts.
We choose as a reference point the $T_{asc,ref}$ value in Table 1 of
\citet{har09}, and we calculate the residuals
$\Delta\,T_{asc}=T_{asc,i}-\left(T_{asc,ref}+N\,P_{b}\right)$, where
$T_{asc,i}$ is the $i-th$ outburst and $N$ is the closest integer to
$\left(T_{asc,i}-T_{asc,ref}\right)/P_b$. The chosen $P_{b}$ is also
reported in Table 1 of \citet{har09}.  

\begin{figure}[t]
  \begin{center}
    \rotatebox{-90}{\includegraphics[width=1.0\columnwidth]{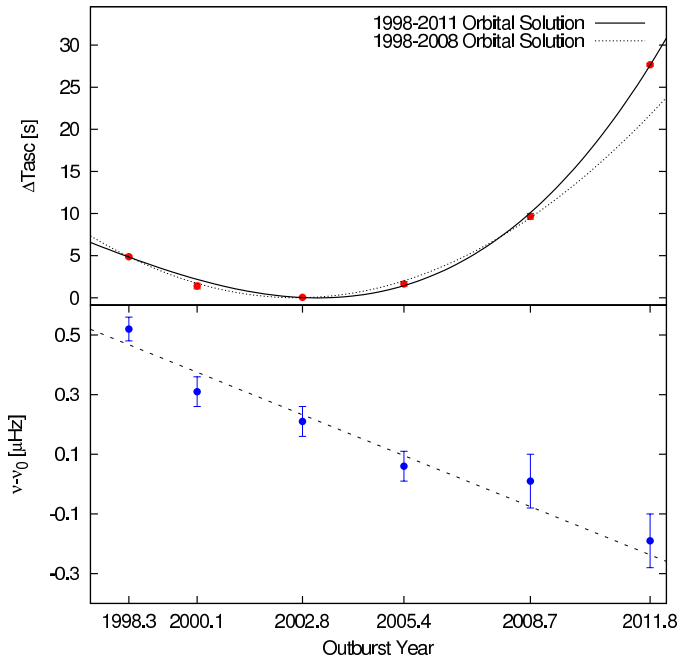}}
  \end{center}
  \caption{Orbital (top panel) and spin frequency evolution (bottom
    panel) over 13 years of observations. The $T_{asc}$ residuals
    cannot be fitted with a parabola (dotted line) and they need a
    cubic fit suggesting an acceleration of the orbital period
    $\ddot{P}_b=1.65\times10^{-20}\rm\,s\,s^{-2}$. The spin-down of
    the pulsar is stable and proceeds at a rate of
    $-10^{-15}\hzs$. The reference frequency is
    $\nu_0=400.975210\rm\,Hz$.\label{fig2}}
\end{figure}

In a previous work \citep{har08,har09} we detected a
$\dot{P}_b=(3.80\pm0.06)\times10^{-12}\rm\,s\,s^{-1}$ (see also
\citealt{dis08} and \citealt{bur09}). When including the 2011 data, a
parabolic fit gives a $\chi^2=62.8$ for 3 dof, thus suggesting that a
constant increase of the orbital period is not the correct model for
\saxj\,. We then add an orbital period second derivative
$\left(\ddot{P}_b\right)$ and fit a cubic polynomial to the data. The
fit is statistically acceptable, with $\chi^2=5.4$ for 2 dof and
parameters $\dot{P}_b=3.5(2)\times10^{-12}\rm\,s\,s^{-1}$ and
$\ddot{P}_b=1.65(35)\times10^{-20}\rm\,s\,s^{-2}$ (see
Figure~\ref{fig2}).

A third time derivative or a sinusoidal model are not required by the data. 
We also tried to fix the $P_b$ and $\dot{P}_b$ at
the values observed up to 2008 and add a sinusoidal fit to the data to
investigate the possibility of apsidal motion. The fit is
statistically unacceptable and the periodicity found is of the order
of $10^3$ yr, which is already orders of magnitude larger than 
 the expected general relativistic effect alone.

\section{Discussion}

\subsection{Pulsar Spin Evolution}


The long term spin-down continues with a constant rate comparable to
what has been measured between 1998 and 2008. The only plausible
explanation for the spin-down in \saxj\ requires magnetic-dipole
radiation with a surface magnetic field of the neutron star at the
poles $B\approx\left(1.5-2.5\right)\times\,10^8\rm\,G$ (for a 
radius $R=10\rm\,km$) for a magnetic-dipole moment
$\mu=\left(0.7-1.5\right)\times10^{26}\rm\,G\,cm^3$. This range
includes the different spin-down measurement methods reported in
\S~\ref{mc} and \S~\ref{phaseflux} and is close to the $B$ field
obtained with Fe line spectral fitting
\citep{cac09,pap09} and accretion disk modeling \citep{ibr09}.

The remarkably constant long term spin-down places stringent
constraints on any ongoing spin-up during an outburst.  If spin
frequency variations were larger than the upper limit reported by
\citet{har08} ($|\nudot|\lesssim 2.5\times10^{-14}\hzs$), they would
produce a scatter in the observed spin frequencies of the order of
$0.1-0.4\uHz$ (for $\nudot\sim5-10\times10^{-14}\hzs$). This scatter
is not observed with statistical errors of $\approx0.05-0.1\uHz$ (see
Fig~\ref{fig2}), and suggests that the net spin-up during an outburst
is at best very small in magnitude for \saxj\,.

\subsection{Orbital Period Evolution}

The orbital period $P_b$ indicates that the orbit of \saxj\ is expanding
and accelerating at a very fast rate. 
The timescale for the acceleration is:
\begin{equation}
\tau_{acc}\cong\frac{2\dot{P}_b}{\ddot{P}_{b}}\sim10\rm\,yr
\end{equation}
suggesting that the identified $\dot{P}_b$ might not represent the
secular evolution of the orbital period. If we assume that the
measured acceleration is constant, then \saxj\ has changed sign of
$\dot{P}_b$ about 25 years ago. There is, however, no reason for
$\ddot{P}_b$ to be constant, since we are not sensitive to higher
order derivatives and both $\dot{P}_b$ and $\ddot{P}_b$ might be part
of long timescale variations similar to those observed in binary
millisecond pulsars \citep{nic00}. Until the 2008 outburst, when only
a constant $\dot{P}_b$ was detected, two interpretations were given: a
secular orbital evolution due to non-conservative mass transfer
(\citealt{dis08}, \citealt{bur09}) and a short-term evolution
associated with exchange of angular momentum between the donor star
and the orbit \citep{har08, har09}.

If \saxj\ has increased its mass loss due to an enhancement of the
donor ablation, then the wind loss from the companion ($\dot{M}_w$)
needs to increase at a rate comparable with that of the orbital
period, since $\dot{P}_b\propto\,\dot{M}_w$. The mass loss is related to
$\dot{E}_{abl}=0.25 \left(R/A\right)^2\dot{E}$, where
$\dot{E}_{abl}$ is the ablation power, $\dot{E}$ is the pulsar
rotational spin-down power and $R$ and $A$ the donor radius and the
semi-major axis of the binary. To explain the acceleration
$\ddot{P}_b$, the energy loss of the pulsar $\dot{E}$ needs to have
increased (in absolute value) in the last 13 years by a factor of
$\approx5$.  Since $\dot{E}\propto \nu\dot{\nu}$, the spin-down
$\nudot$ needs to vary at a rate $\ddot{\nu}\simeq
-10^{-23}\rm\,Hz\,s^{-2}$ to reach the energy loss required. By using
the long-term spin evolution presented in the previous section we can
put constraints on $|\ddot{\nu}|\lesssim10^{-24}\rm\,Hz\,s^{-2}$.  We
conclude that the enhanced ablation scenario is not supported by the
observations. 

A dynamically induced period derivative in the gravitational potential
well of a third body can also be excluded. The effect of a potential
well is identical on the orbital and spin frequencies and derivatives:
\begin{equation}
f^{(n)}=-f\frac{\mathbf{a}^{(n-1)}\cdot\,\mathbf{\hat{n}}}{c}
\end{equation}
where $f^{(n)}$ is the n-th time derivative of the orbital or spin
frequency, $\mathbf{a}$ is the acceleration due to the third body,
$\mathbf{\hat{n}}$ is a unit vector along the line of sight and $c$
the speed of light. To explain the observed $\ddot{P}_b$ we need
$\dot{\mathbf{a}}\sim10^{-15}-10^{-16}\rm\,m\,s^{-3}$
and $|\ddot{\nu}|\sim10^{-21}\hzs$, which is not observed.

If the measured orbital evolution is a short-term event, 
then one explanation can be found with the donor
spin-orbit coupling model. A coupling between the pulsar rotational energy
loss (in form of winds or fields) and the orbital angular momentum
\citep{dam91} is ruled out by the small magnitude of the effect
produced by the tiny $\dot{E}$ of \saxj\,.
A mass quadrupole variation of the donor star is a more promising
possibility. 
A change $\Delta Q$ in the mass quadrupole leads to a change in orbital 
period \citep{ric94, app94, app87}:
\begin{equation}
\frac{\Delta P_b}{P_b}=-2\left(\frac{R}{A}\right)^2\frac{\Omega^2 R^3}{G M}\frac{M_s}{M}\frac{\Delta\Omega}{\Omega}
\end{equation}
where $M$ and $R$ are the donor mass and radius, $M_s$ is a thin shell
of mass generating the quadrupole, 
and $\Omega$ the angular velocity of the star.  If we assume that the angular
velocity of the donor is almost synchronous with the orbital angular
velocity, then the variation
$\Delta P_b\simeq 0.004\rm\,s$ observed in the last 13 years gives:
\begin{equation}
\frac{\Delta \Omega}{\Omega}\sim 10^{-4}\frac{M}{M_s}
\end{equation}

The observed orbital period variations in the eclipsing millisecond
pulsar PSR J2051-0827 and PSR B1957+20 are likely to be caused by
changes in the quadrupole moment of the companion
\citep{arz94,dor01,laz11}.
\citet{app92} proposed a magnetic activity cycle that leads to a
deformation of the star at the origin of this behavior. The donor star
of \saxj\ is also in Roche lobe
contact, whereas binary millisecond pulsars are detached systems. If
the orbital period of \saxj\ has decreased in the past for some time,
then the Roche lobe has moved across the outer envelope of the
brown dwarf enhancing the mass transfer rate.
A detailed discussion of this effect is beyond the scope of this
letter, but we can speculate that \saxj\ has gone through periodic
episodes (each lasting $\tau_{acc}\sim10$ yr) of enhanced accretion in
the past. This effect is opposite during the accelerated 
orbital expansion, with the mass transfer being less than 
in the non-accelerated case. The quadrupolar moment
change has also the effect of heating the star, providing an
explanation for the large entropy content of the donor \citep{del08}.



\bigskip

\acknowledgements{AP acknowledges support from an NWO-Veni fellowship. RW
  was partly supported by an ERC starting grant}


\end{document}